\documentclass[aps,prl,reprint,showkeys,floatfix]{revtex4-1}
\usepackage{graphicx}
\usepackage[utf8]{inputenc}
\usepackage{amsmath}
\usepackage{verbatim}
\usepackage{amssymb}
\usepackage{listings}
\usepackage{color}
\renewcommand\vec[1]{\boldsymbol{#1}}

\newcommand\Fig[1]{Fig.~\ref{#1}}

\newcommand\quot[1]{``#1''}

\newcommand{\PCAL}{\mathcal{P}} 

\newcommand{\XYM}{$XY$ model}
\newcommand{\XYSG}{$XY$ spin glass model}
\newcommand\eq[1]{Eq.~\ref{#1}}
\newcommand{\lift}[1]{\stackrel{\curvearrowright}{_#1}}
\begin{document}
\date{\today}\title{Event-chain Monte Carlo for classical continuous spin
models}

\author{Manon Michel}
\email{manon.michel@ens.fr}
\author{Johannes Mayer}
\author{Werner Krauth}
\email{werner.krauth@ens.fr}
\affiliation{Laboratoire de Physique Statistique, Ecole Normale Sup\'{e}rieure
/ PSL Research University, UPMC, Universit\'{e} Paris Diderot, CNRS, 24 rue
Lhomond, 75005 Paris, France}

\begin{abstract}
We apply the event-chain Monte Carlo algorithm to classical continuum spin
models on a lattice and clarify the condition for its validity. In the
two-dimensional \XYM, it outperforms the local Monte Carlo algorithm by two
orders of magnitude, although it remains slower than the Wolff cluster
algorithm. In the three-dimensional \XYSG\ at low temperature, the
event-chain algorithm is far superior to the other algorithms.
\end{abstract}
\maketitle
\section{Introduction}
\label{s:Introduction}
Classical and quantum spin models are of fundamental interest in statistical and
condensed-matter physics. Spin models are also a crucial test bed for
computational algorithms.

An important representative is the model of continuous
two-dimensional classical spins
of fixed length (rotators) on a two-dimensional lattice. Thirty years ago, the
existence and nature of the phase transition in this two-dimensional \XYM\ were
highly controversial\cite{Seiler_1988}. The substitution of the traditional
local Monte Carlo (LMC) algorithm\cite{Metropolis_1953} by Wolff's spin flip
cluster (SFC) algorithm\cite{Wolff_1989} then quickly allowed to clarify
that this model indeed undergoes a Kosterlitz-Thouless
transition\cite{Janke_1993,Hasenbusch_1994}, whose
temperature is now known to five significant digits
\cite{Hasenbusch_2005, Komura_2012}. SFC has played a
decisive role in understanding the physics of the \XYM \cite{KT, Jose_1977,
Amit_1980}, and in arriving at its detailed quantitative description.

SFC and its variants can be implemented for a wide range of models, but they
are efficient only in a few of them. Particularly frustrating is the case
of the three-dimensional \XYSG, where the algorithm loses all its
power\cite{Kawamura_2001, Obuchi_2013}. For this much studied spin glass model,
our understanding today resembles the one of the \XYM\ before the revolution
triggered by the cluster algorithms. Clearly, there still is a great need for
more powerful algorithms for classical and quantum spin models.

Today's Markov-chain Monte Carlo algorithms generally follow the conventional
paradigm based on three principles: 1/ Each move represents a finite change of
the configuration. It is independent of the previous move, and depends only on
the configuration itself. 2/ The algorithm satisfies the detailed-balance
condition. 3/ The decision whether a proposed move is accepted is based
on the change in energy, using the Metropolis acceptance rule or the
heat-bath condition\cite{Metropolis_1953,SMAC}.

In the present work, we show that the novel event-chain Monte Carlo
(ECMC) paradigm\cite{Bernard_2009, Michel_2014,Peters_2012}, that has
already been very successful in particle systems
\cite{Bernard_2011,Isobe_2015,Kapfer_2015,Kampmann_2015}, can also be
applied to the \XYM\ and the \XYSG. The paradigm breaks all three
principles of the conventional Markov-chain scheme: Moves are
infinitesimal rather than finite, although an event-driven
scheme allows to recover finite
displacements\cite{Peters_2012}. In one-dimensional systems, 
the moves do not change with time. In multidimensional
systems, moves persist on long time scales. This is achieved within the
Markov-chain scheme through additional \quot{lifting}
variables\cite{Diaconis_2000,Michel_2014}. In addition, ECMC
violates detailed balance and only satisfies the weaker global
balance condition (cf.
\cite{Turitsyn_2011,Suwa_2011,Fernandes_2011,Sakai_2013,Ichiki_2013}). Finally,
the decision on future moves is based on the change in pair energies, rather
than the change in total energy. This is achieved by replacing the standard
Metropolis algorithm by its recently introduced factorized
variant\cite{Michel_2014}.

For the two-dimensional \XYM\ at the critical point, we find that
ECMC is about 100 times faster than LMC, although the
presence of a slow time scale in autocorrelation functions makes that it is not
as fast as SFC. In the low-temperature phase of the three-dimensional \XYSG,
where SFC is known to be inefficient, ECMC clearly outperforms LMC.

\section{From local Monte Carlo to the \quot{Event-chain} algorithm}
\label{s:ecmc}
\begin{figure}[ht]
\begin{center}
\includegraphics[width=\columnwidth]{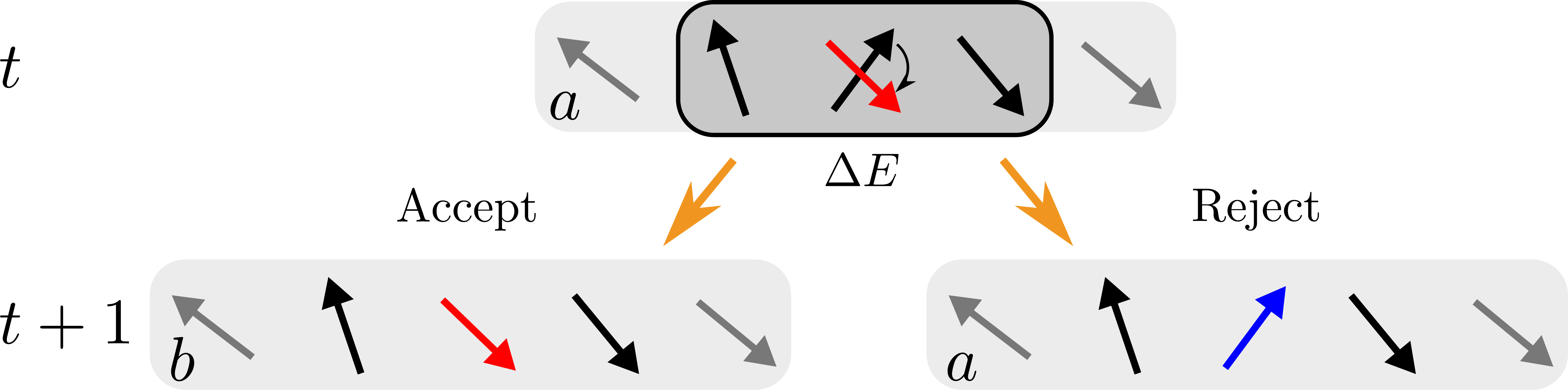}
\caption{LMC move for the one-dimensional \XYM.
\emph{Upper panel}: Configuration at time $t$ and
proposed displacement $\Delta \phi$ of a randomly chosen spin,
corresponding to an energy change $\Delta E$. \emph{Lower panel}:
Possible configurations at time $t+1$: The proposed move is accepted with
probability $\min(1, \exp(-\beta \Delta E)$ (\emph{left}) and
rejected otherwise (\emph{right}).}
\label{f:Metropolis_standard}
\end{center}
\end{figure}
In the two-dimensional ferromagnetic \XYM\ of spins $\vec S_k = (S_k^x,S_k^y) =
(\cos\phi_k,\sin\phi_k)$ on a lattice with sites $i=1, \ldots, N$, and with
an energy
\begin{equation}
E = - \sum_{ \langle i, j\rangle} J_{ij}\vec S_i\cdot \vec S_j =
\sum_{ \langle i, j\rangle}
\underbrace{\left[ - J_{ij} \cos(\phi_i-\phi_j) \right]}_{E_{ij}},
\label{e:energy_XY}
\end{equation}
the coupling constants $J_{ij}$ are all equal to one. The sum $\langle i, j
\rangle$ goes over nearest neighbors on the lattice. We refer to
the $ E_{ij}$ as \quot{pair energies}. The \XYM\ on a two-dimensional square
lattice undergoes a phase transition at inverse temperature $\beta = 1.1199$,
see ref. \cite{Hasenbusch_2005}.

In LMC, one proposes at each time step $t$ a finite move from a
configuration $a$ to a configuration $b$ (a rotation by a finite angle $\Delta
\phi$ of a spin $k$), as sketched in \Fig{f:Metropolis_standard}. To satisfy
detailed balance\cite{SMAC}, $k$ is randomly chosen at each time step,
and $\Delta \phi$
is sampled from a symmetric distribution around zero, so that $\Delta \phi$
arises with the same probability as $-\Delta \phi$. The proposed move
corresponds to an energy change $\Delta E = E_b - E_a$ in \eq{e:energy_XY}, and
it is accepted with probability
\begin{equation}
p_{\text{acc}}^{\text{Met}} = \min(1, \exp(-\beta \Delta E)).
\label{e:metropolis_filter}
\end{equation}
The exponential in this equation corresponds to the ratio $\pi_b / \pi_a$ of the
Boltzmann weights of the configurations.

Practically, the move is accepted, and the configuration updated to $b$, if a
uniform random number between $0$ and $1$ satisfies
$\text{ran}(0,1) < p_{\text{acc}}^{\text{Met}}$ (see \cite{SMAC}). Otherwise,
the
configuration at time $t+1$ is the same as the one at time $t$, namely $a$.

\begin{figure}
\begin{center}
\includegraphics[width=\columnwidth]{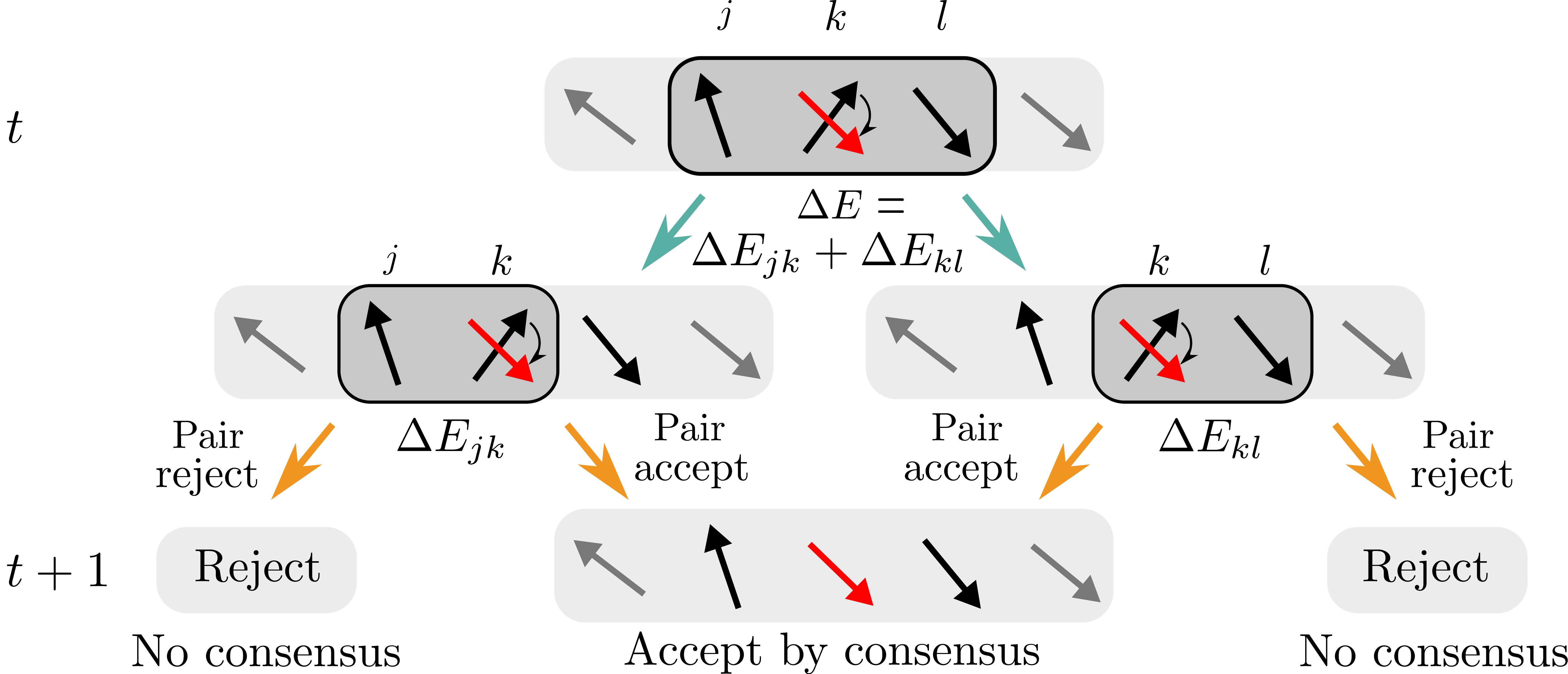}
\caption{Factorized Metropolis move. \emph{Upper panel}: Configuration at time
$t$ and proposed displacement $\Delta\phi$ of a randomly chosen spin $k$.
\emph{Middle panel}: Factorization into pairs $(j,k)$ and $(k,l)$. In the
factor $(j,k)$, the move is pair-accepted with probability $\min(1,
\exp(-\beta \Delta E_{jk})$, etc. \emph{Lower panel}: Possible configurations
at time $t+1$: The proposed move is either accepted by consensus (i.e.
independently by all pairs) or else rejected.}
\label{f:Metropolis_factorized}
\end{center}
\end{figure}

The recently introduced factorized algorithm\cite{Michel_2014} also satisfies
the detailed-balance condition. In this method, the energy-based Metropolis
acceptance probability is replaced by a factorized form which separately depends
on the pair-energy changes:
\begin{equation}
p_{\text{acc}}^{\text{fact}} = \prod_{\langle k,l \rangle}
p_{\text{acc}}^{kl} =
\prod_{\langle k,l \rangle}
\min(1, \exp(-\beta \Delta E_{kl})).
\label{e:factor_filter}
\end{equation}
The proposed move $a \to b$ is accepted with this probability. The
factorized algorithm always has a smaller acceptance rate than the
conventional one, $p_{\text{acc}}^{\text{fact}} \le
p_{\text{acc}}^{\text{Met}}$ (this will however turn out not to be a problem
in ECMC). To implement \eq{e:factor_filter}, one might use a single random
number and accept the move if $\text{ran}(0,1) < p_{\text{acc}}^{\text{fact}}$.
We rather accept the move if several independent random numbers
satisfy $\text{ran}_{kl}(0,1) <
p_{\text{acc}}^{kl}$ for all pairs $k,l$. In other words, a move is
accepted only if it is pair-accepted by all pairs $k,l$.
This consensus rule is illustrated in \Fig{f:Metropolis_factorized}.
We note that the factorization in
\eq{e:factor_filter} relies on the possibility to cut the hamiltonian into
independent pieces. The factorization
may also be used to separate different components
of the inter-particle potential, as for example the $1/r^6$ and
$1/r^{12}$ pieces in the Lennard-Jones potential
\cite{Michel_2014,Kapfer_2015}.

\begin{figure}
\begin{center}
\includegraphics[width=0.9 \columnwidth]
{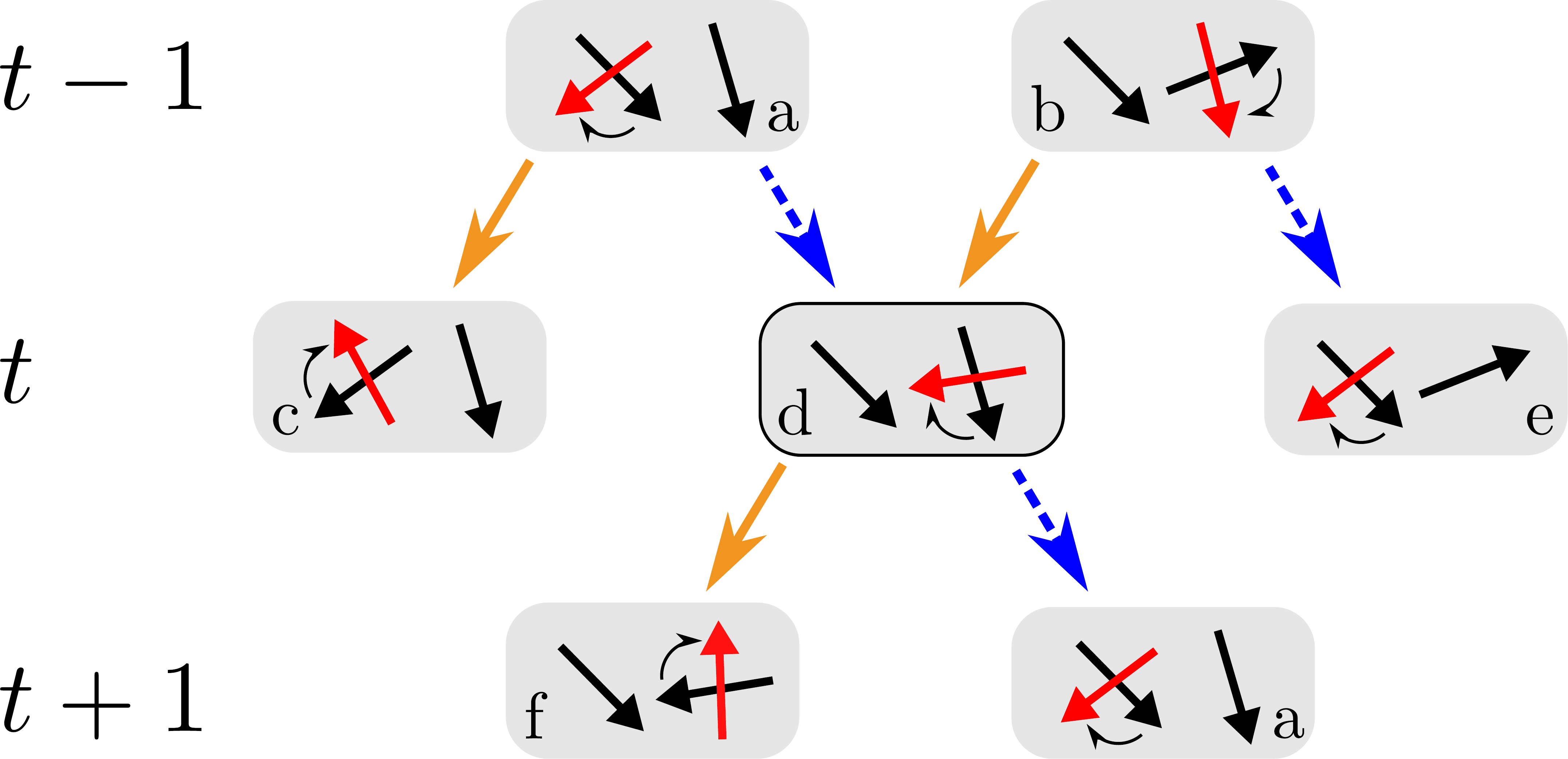}
\caption{Lifting approach of ECMC. Physical moves $b \to d$, $d \to f$ and
$a \to c$ are by the same infinitesimal angle $\Delta
\phi$ in clockwise direction, all others are lifting moves that
preserve the physical configuration. Note that $\pi_b = \pi_c$ because of
\eq{e:energy_XY}.}
\label{f:Metropolis_infinitesimal}
\end{center}
\end{figure}

The ECMC combines the factorized Metropolis probability with
the \quot{lifting} concept of Diaconis et al.\cite{Diaconis_2000} and
with the idea of infinitesimal displacements\cite{Michel_2014}. The term
\quot{lifting} refers to the extension of the physical configuration by an
additional variable that fixes the proposed move. Written as $\lift{k}$, it
singles out the spin $k $ as the only one that can move, as $\phi_k \to \phi_k +
\Delta \phi$ (see \Fig{f:Metropolis_infinitesimal}). If the move is accepted,
the lifting variable for the next time step $t+1$ is again $\lift{k}$.
If the physical move is rejected, a lifting move takes place and the lifting
variable is passed on to the spin $l$ of the pair that rejected the move,
and the physical configuration is unchanged. In both cases, the value of $\Delta
\phi$ is used again.
Note that for infinitesimal $\Delta
\phi$, the acceptance probabilities of the physical moves approach one and the
rejection probabilities approach zero. Multiple rejections are totally
suppressed, and  the choice of $\lift{l}$ is unique\cite{Michel_2014}. At
each time step, either a lifting move or a physical move takes place, and
ECMC is thus formally rejection-free.

ECMC satisfies the global balance condition in the \XYM, as we now show:
For simplicity, we consider only two spins and concentrate on a configuration
$d$ (see \Fig{f:Metropolis_infinitesimal}). This configuration
can only be reached through a lifting move from $a$ or
through a physical move from $b$. The global-balance
condition\cite{SMAC} states that the flow into configuration $d$ must be
equal to the flow out of it:
\begin{multline}
\underbrace{\pi_a p (a \to d)}_{\PCAL(a \to d)} +
\underbrace{\pi_b p (b \to d)}_{\PCAL(b \to d)}
=\\
\underbrace{\pi_d p (d \to f)}_{\PCAL(d \to f)} +
\underbrace{\pi_d p (d \to a)}_{\PCAL(d \to a)}.
\label{e:global_balance}
\end{multline}
Here, $\PCAL(a \to d)$ represents the probability flow from $a $ to $d$, etc.
For ECMC, the probabilities $p$ in \eq{e:global_balance}
coincide with the acceptance probabilities: All configurations carry
a lifting variable that specifies the spin that may move 
and the move itself, $\Delta \phi$.

The statistical weight $\pi_a$ is trivially equal to $\pi_d$ because they
differ only by a lifting move. Furthermore, $\pi_c$ equals $\pi_b$, as the two
configurations differ only by a global rotation. Writing $\Delta E = E_b - E_d$,
we thus find
\begin{multline}
\PCAL(b \to d) = \pi_b p_{\text acc}^{\text{frac}}(b \to d) = \pi_d p_{\text
acc}^{\text{frac}}(d \to b) \\
= \pi_d \min(1, \exp(- \beta \Delta E)).
\end{multline}
Note in this equation that $\pi_b p_{\text acc}^{\text{frac}}(b \to d) = \pi_d
p_{\text acc}^{\text{frac}}(d \to b)$, because the factorized transition
probabilities satisfy detailed balance. Likewise, the change in energy in
going from $a \to c$ is also $\Delta E$ and $p(a \to d) = 1 - p(a \to c)$.
Therefore, the flow $\PCAL(a\to d)$ satisfies
\begin{multline}
\PCAL(a \to d) = \pi_a ( 1 - \min(1, \exp(- \beta \Delta E)) \\
= \pi_d ( 1 - \min(1, \exp(- \beta \Delta E)).
\end{multline}
It follows that the flow into $d$, namely the sum of $P(a \to d)$ and of $P(b
\to d)$, equals $\pi_d$. As for the flow out of $d$, it trivially equals
$\pi_d$ because of the conservation of probabilities. It follows that the global
balance of \eq{e:global_balance} is satisfied. The factorization property and
the infinitesimal limit guarantee that the argument carries over to general $N$
(see \cite{Michel_2014}).

ECMC violates the detailed balance condition $\PCAL(b \to d) = \PCAL(d \to
b)$: A move $d \to b$ would be anti-clockwise, yet all moves within ECMC
are, by the initial choice of $\Delta \phi$, clockwise. Also, $\PCAL(a \to d) =
0$, as $E_d > E_f$ and all physical moves
from $d$ to $f$ are accepted. Furthermore, for ECMC to be valid, the pair
energy must be symmetric (so that $\pi_b = \pi_c$ in
\Fig{f:Metropolis_infinitesimal}). Modified \XYM s, as described in
ref.\cite{Domany_1984}, can also be treated, but more general pair energies
require special considerations\cite{Michel_2015}.

ECMC with infinitesimal moves requires a scaling of physical time:
In one unit of time, as $\Delta \phi$ goes to zero, an infinite number
of physical moves take place, but the number of lifting moves remains
finite. In an event-driven approach\cite{Peters_2012,Michel_2014}, the
algorithmic complexity can be made to scale with the number of liftings: The
lifting variable being set to $\lift{k}$, the angle $\phi_k$ now rotates
clockwise until the \quot{event}, i.e. a lifting move, is produced through a
rejection by a neighbor $l$. The lifting variable is updated to $\lift{l}$,
$\phi_l$ rotates clockwise, etc. Effectively, one undergoes an infinite number
of Monte Carlo steps, giving a continuous trajectory.

\begin{figure}
\begin{center}
\includegraphics[width=0.9 \columnwidth]
{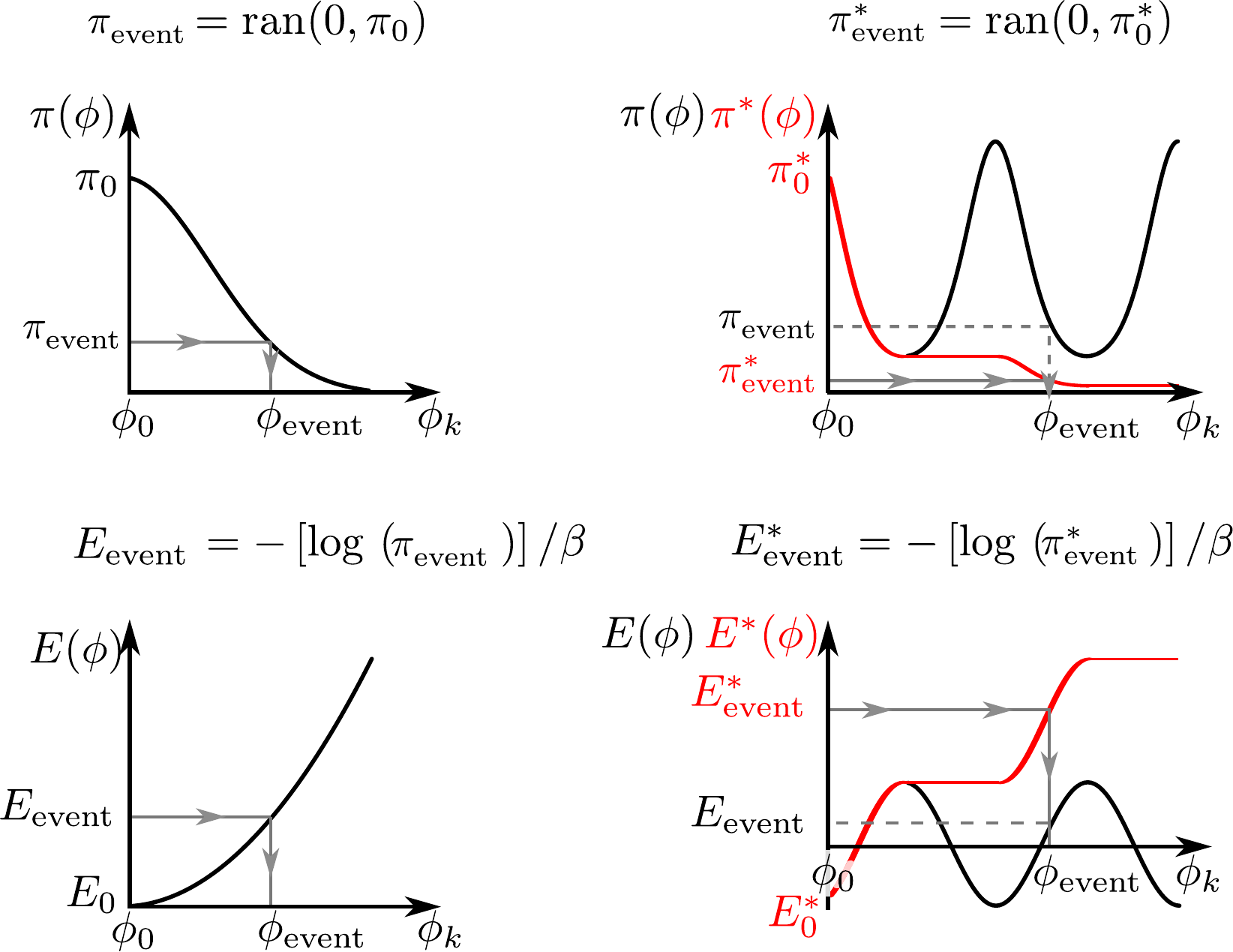}
\caption{Event-driven implementation of ECMC for a pair of spins
$(k,l)$. From a starting point $\phi_k = \phi_0$ of weight
$\pi_0$ and energy $E_0$, $\phi_k$ is updated by infinitesimal
moves until $\phi_k = \phi_{\text{event}}$. 
\emph{Left:} Monotonously decreasing distribution $\pi$: The lifting event is
sampled as $\pi_{\text{event}} = \text{ran}(0, \pi_0)$.
\emph{Right:} General distribution $\pi$:
$E^*_{\text{event}} - E^*(0) = [- \log\  \text{ran}(0, 1)] / \beta$.
}
\label{f:sampling_pi_E}
\end{center}
\end{figure}

The angle $\phi_k$ corresponding to the next event is easily sampled:
We continue to consider a single
pair $(k,l)$ of spins, with the lifting variable $\lift{k}$. The $i$-th
infinitesimal update of $\phi_k$ is noted as the move $i-1 \to i$ and the weight
of the configuration $(\phi_i = \phi_k + i d\phi, \phi_l)$, $\pi_i$. The
probability $p_{\text{event}}(0 \to n)$ to accept $n$ subsequent physical moves
and then to reject the $n + 1$st physical move is
\begin{multline}
p_{\text{event}}(0 \to n) =
p_{\text{acc}}(0 \to 1) \cdots
p_{\text{acc}}(n-1 \to n)\\
\left[ 1 - p_{\text{acc}}(n \to n + 1) \right].
\label{e:event_prob}
\end{multline}
The $j$th term in this expression is $\min(1, \pi_j/\pi_{j-1}$).
Supposing for a moment that $\pi_j$ is monotonously
decreasing with $j$, this gives
\begin{equation}
\begin{aligned}
p_{\text{event}}(0 \to n) &= \frac{\pi_{n-1}}{\pi_0}\left(1 - \frac{\pi_n}{\pi_{n-1}} \right)\\
&= \frac{-1}{\pi_0}
\left.\frac{\partial \pi}{\partial \phi_k}\right|_{\phi_k = \phi_n} d \phi.
\label{e:p_rej_neg}
\end{aligned}
\end{equation}
This probability is normalized, writing $\phi_{\text{event}}$ the value of
$\phi_k$ at which the event happens:
\begin{multline}
- \frac{1}{\pi_0} \int_{0}^{\infty}\left. \frac{\partial \pi}{\partial
\phi_k}\right|_{\phi_k = \phi_{\text{event}}} d\phi_{\text{event}}\\
= 
\frac{1}{\pi_0} \int_{0}^{\pi_0} d \pi_{\text{event}} = 1.
\end{multline}
This integral is sampled by \cite{SMAC}
\begin{equation}
\begin{aligned}
\pi_{\text{event}} &= \text{ran}(0, \pi_0)\\
\pi_{\text{event}}/\pi_0& = \text{ran}(0, 1),
\label{e:random_uniform_in_probability}
\end{aligned}
\end{equation}
which is equivalent to the following sampling of the energy increase:
\begin{equation}
\Delta E(\phi_{\text{event}}) = - \left[\log\ \text{ran}(0, 1)\right]/ \beta.
\label{e:random_uniform_in_energy}
\end{equation}
Sampling $\pi$ uniformly between
$0$ and the present value, $\pi_0$ (equivalently, $\Delta E$ from its
exponential distribution) thus yields the event time, $\phi_{\text{event}}$ (see
\Fig{f:sampling_pi_E}).

For a non-monotonous probability distribution, all negative energy increments
correspond to an acceptance probability $1$, and disappear from
\eq{e:event_prob}. The sampling of the energy increase in
\eq{e:random_uniform_in_energy} turns into the sampling of only the
positive energy changes. 
As shown in \Fig{f:sampling_pi_E}, this can be expressed as a function
$E^*$, constructed only from the positive increments of the energy $E$
\cite{Peters_2012}.

For a system of more than one pair of spins, the event times
$\phi_{\text{event}}$ for each neighbor of the lifted spin $k$ can be computed
independently in view of the factorized probability of \eq{e:factor_filter}, and
$k$ turns clockwise up to the earliest event (that involves, say,
another spin $l$). The lifting variable is then set to $\lift{l}$.

It follows from \eq{e:event_prob} that all configurations encountered between
two events sample the Boltzmann distribution. Any uniform subset of these
configurations can be used for averaging observables. A practical choice
consists in outputting spin configurations at regular intervals independent of
the occurrence of events.

For the models considered here, we found that the efficiency was not increased
by halting and restarting the simulation after fixed displacements. In contrast,
switches between moves along the different coordinate axes assure ergodicity in
multi-dimensional hamiltonians as they appear in particles
systems\cite{Bernard_2009}, but also the related Heisenberg
model\cite{Hukushima_2015}.

\section{Simulations for the two-dimensional \XYM\ at the critical point}
\begin{figure*}[htb]
\begin{center}
\includegraphics[width= 2 \columnwidth]
{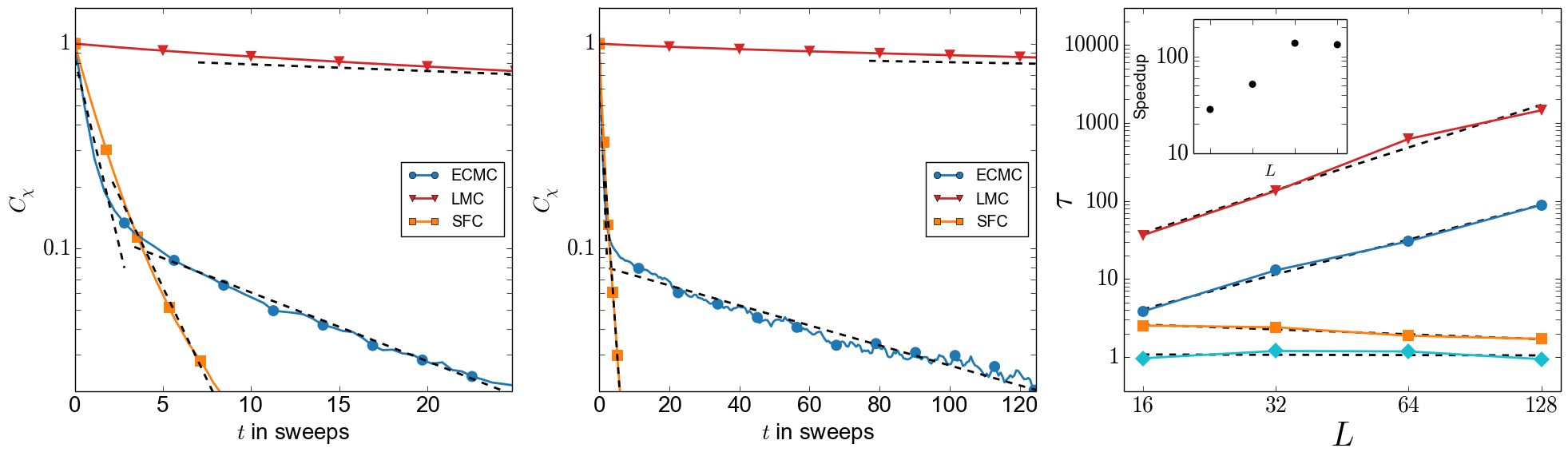}
\caption{Autocorrelation function $C_{\chi}(t)$ for the
two-dimensional \XYM\ at the critical point $\beta = 1.1199$ for
LMC (\emph{red, triangle}), ECMC (\emph{blue, circle}), and
SFC (\emph{yellow, square}). Exponential fits (\emph{black, dotted}) are as in
\eq{e:exponential_fits}. \emph{Left:} $N = 32^2$.
\emph{Middle:}
$N = 128^2$. \emph{Right:} Scaling of the
autocorrelation time $\tau$ with the system size. Both LMC (\emph{red,
triangle}) and the slow scale of ECMC
(\emph{dark blue, circle}) are compatible with a dynamical scaling exponent $z
\sim 2$. Both the fast scale of ECMC (\emph{light blue,
diamond}) and SFC (\emph{yellow, square}) are compatible with $z \sim 0$.
\emph{Right Inset:} Speedup of ECMC with respect to LMC \emph{vs.}
$L$.}
\label{f:XY_speedup}
\end{center}
\end{figure*}
In the two-dimensional \XYM, we consider the susceptibility $\chi$
\begin{equation}
\chi = \frac{|| \sum \vec S_k ||^2}{N},
\end{equation}
and estimate the convergence properties by the susceptibility autocorrelation
function
\begin{equation}
C_{\chi}(t) = \frac{\langle \chi(t' + t) \chi(t') \rangle - \langle \chi \rangle^2}
                  {\langle \chi^2 \rangle - \langle \chi \rangle^2} 
\end{equation}
at the critical point $\beta= 1.1199$ (see \cite{Hasenbusch_2005}). We suppose
that
$\chi$ is a slow variable of this model.
We measure time in sweeps: For ECMC, one sweep corresponds to $\sim N$
lifting events while for LMC, one sweep corresponds to $ N$
attempted moves. For SFC, a sweep denotes $\sim N$ spins added to
clusters. The complexity of one sweep is $O(N)$ in the three algorithms
and the CPU times used per sweep are roughly comparable.

In \Fig{f:XY_speedup}, we show the autocorrelation function for the \XYM\ at its
critical point, obtained from very long single runs of the
algorithms. For LMC and SFC, the decay of the susceptibility
autocorrelation function can be described by a single time scale, while for
ECMC, it is well described by two time scales:
\begin{equation}
C_{\chi}(t) \simeq
\begin{cases}
\exp(- t / \tau^\text{LMC})\quad \text{(LMC)} \\
\exp(- t / \tau^{\text{SFC}})\quad \text{(SFC)} \\
A_0 \exp(- t / \tau^{\text{ECMC}}_0) + \\ A_1 \exp( - t /
\tau^{\text{ECMC}}_1)\quad \text{(ECMC)}
\end{cases} .
\label{e:exponential_fits}
\end{equation}
For ECMC, this correlation function rapidly decays to $C_\chi \sim 0.1$ on a
timescale $\tau_0^{\text{ECMC}}$ of about 5 sweeps. A slow
mode $\tau_1^{\text{ECMC}}$ then sets in. It presents a $z=2$ scaling
($\tau_1^{\text{ECMC}} \sim L^2$, with $N = L^2$). As shown on the right
panel of \Fig{f:XY_speedup}, $\tau_1^{\text{ECMC}}$ is an order of magnitude
smaller than $\tau^{\text{LMC}}$. Together with the initial rapid decrease,
this
makes ECMC about one hundred times faster than LMC. However, its
dynamical scaling exponent appears to be $z \sim 2$, as for LMC.
We notice that in particle systems, ECMC also shows initial ballistic behavior,
but then crosses over into slower decay\cite{Kapfer_2013}.
\section{Three-dimensional \XYSG}
\begin{figure}[htb]
\begin{center}
\includegraphics[width=0.8\columnwidth]
{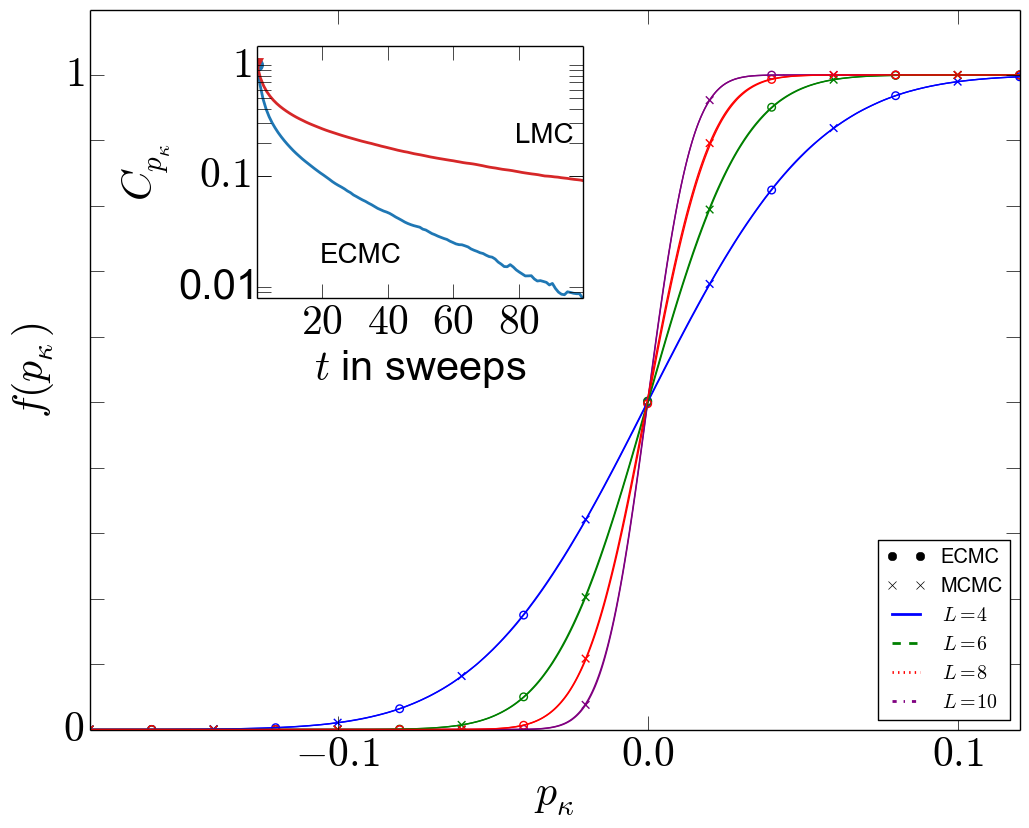}
\caption{Cumulative distribution of the chiral overlap $p_\kappa$ for the
three-dimensional \XYSG\ for $N = 4^3, 6^3, 8^3, 10^3$ at $\beta = 1.5$, in the
high-temperature phase (single samples). \emph{Inset}: Autocorrelation
function $C_{p_\kappa}(t)$ for $N = 6^3$ from LMC (\emph{red, triangle}) and
ECMC (\emph{blue,
circle}). }

\label{f:cumul_SG}
\end{center}
\end{figure}

We now study ECMC for the three-dimensional \XYSG, where the
nearest-neighbor coupling constants $J_{ij}$ are drawn from a Gaussian normal
distribution of zero mean and unit variance. The algorithm can be formulated
as for the ferromagnetic model,  and the spins continue to always turn 
clockwise. We will find evidence that the relaxation dynamics of ECMC
differs from the one of LMC. Following \cite{Kawamura_2001}, we consider the
chiral overlap between two independent systems, $(1)$ and $(2)$, with identical
coupling constants
\begin{equation}
p_\kappa = \frac 1N \sum^N_{p=1}\kappa^{(1)}_{p \perp \mu}\kappa^{(2)}_{p
\perp \mu},
\end{equation}
with $\kappa^{(i)}_{p \perp \mu}$ being the chirality at a plaquette
$p$, perpendicular to the axis $\mu$, defined as:
\begin{equation}
\kappa^{(i)}_{p \perp \mu} = \frac{1}{2 \sqrt 2} \sum_{(i,j) \in p} \text{sgn}
(J_{ij}) \sin(\phi_i-\phi_j).
\end{equation}
The sum $\sum_{(i,j) \in p}$ is taken over the four bonds encircling
the plaquette $p$ clockwise. By construction,
$p_\kappa$ is a symmetric function about zero. As shown in
\Fig{f:cumul_SG}, ECMC and LMC agree very well at high temperature.
The autocorrelation function of the chiral overlap for LMC and
ECMC, shown in \Fig{f:cumul_SG}, gives at high temperature a
size-independent speedup by a factor $\sim 5$ of ECMC. 

\begin{figure}[htb]
\begin{center}
\includegraphics[width=0.8\columnwidth]
{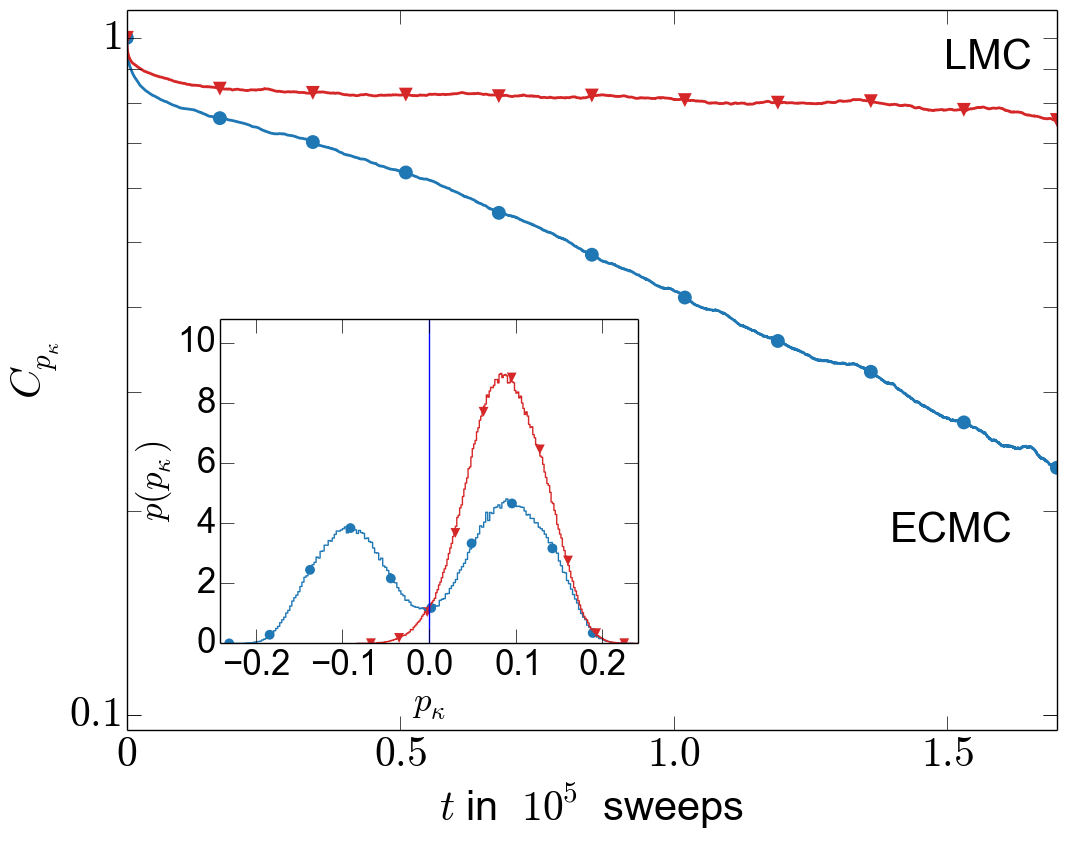}
\caption{Chiral overlap autocorrelation function from ECMC and LMC at 
$\beta = 3.636$ for a given sample at $N = 6^3$.
\emph{Inset}: Distributions of $p_\kappa$ after $10^6$ sweeps for the two
algorithms in the same sample. Note the nearly symmetric distribution for ECMC.}
\label{f:speedup_SG}
\end{center}
\end{figure}

The phase diagram of the three-dimensional \XYSG\ at low temperature
(with the possible existence of separate spin-glass and chiral-glass phases) 
is still being debated. We consider $\beta = 3.636$, which may be the locus of 
the spin glass transition \cite{Obuchi_2013}, or below it, near the transition
\cite{Pixley_2008, Roma_2014}. At this temperature, ECMC exhibits a striking
advantage over LMC in one third of samples of size $N = 6^3$, where it
explores the configuration space more easily, without using parallel tempering
\cite{Hukushima_1996}. A typical example of a symmetric chiral overlap
distribution profile after $10^6$ sweeps (symmetric for ECMC, but not for LMC is
shown in \Fig{f:speedup_SG}, together with the corresponding autocorrelation
function. For larger systems, the speedup of ECMC with respect to LMC seems to
increase, but already for $10^3$ systems, ECMC no longer equilibrates at $\beta
= 3.636$.

\section{Conclusion}
In conclusion, we have applied in this work the recent event-chain algorithm to
classical spin models, and obtained a considerable algorithmic speed-up with
respect to the local Monte Carlo algorithm for the two-dimensional \XYM\ at its
critical point. The new method appears very general, as we also obtained clear
acceleration for the three-dimensional \XYSG\ at low temperature. It will be
interesting to see how well the event-chain algorithm couples with the
traditional acceleration methods, as for example the parallel tempering method,
or the overrelaxation approaches that have been much used for spin glasses.

\section{Acknowledgments}
We thank K. Hukushima for discussion, and for sharing information about
closely related work in his group on the three-dimensional Heisenberg
model\cite{Hukushima_2015}. 

This work was granted access to the HPC resources of
MesoPSL financed by the Region Ile de France and the project Equip@Meso
(reference ANR-10-EQPX-29-01) of the programme Investissements d’Avenir
supervised by the Agence Nationale pour la Recherche.

\end{document}